\documentclass[sigconf, authorversion]{acmart}
\usepackage{multirow, makecell}
\usepackage{tabularx,booktabs}

\AtBeginDocument{%
  }

\copyrightyear{2023} 
\acmYear{2023} 
\setcopyright{acmlicensed}\acmConference[IMC '23]{Proceedings of the 2023 ACM Internet Measurement Conference}{October 24--26, 2023}{Montreal, QC, Canada}
\acmBooktitle{Proceedings of the 2023 ACM Internet Measurement Conference (IMC '23), October 24--26, 2023, Montreal, QC, Canada}
\acmPrice{15.00}
\acmDOI{10.1145/3618257.3624844}
\acmISBN{979-8-4007-0382-9/23/10}

\newcommand{\ie}{i.e., }
\newcommand{\eg}{e.g., }
\newcommand{\cf}{cf. }
\makeatletter
\newcommand{\etal}{~et~al\@ifnextchar.{}{.\@}}
\newcommand{\etc}{etc\@ifnextchar.{}{.\@}}
\makeatother

\newcommand{\afblock}[1]{\noindent{\textbf{#1}}}
\newcommand{\takeaway}[1]{\noindent{\textbf{Takeaway.}} \textit{#1}}

\usepackage{prettyref}
\newrefformat{chap}{Chapter~\ref{#1}}
\newrefformat{app}{Appendix~\ref{#1}}
\newrefformat{sec}{Sec.\,\ref{#1}}
\newrefformat{sub}{Sec.\,\ref{#1}}
\newrefformat{subsec}{Sec.\,\ref{#1}}
\newrefformat{ssub}{Sec.\,\ref{#1}}
\newrefformat{eq}{Eq.~(\ref{#1})}
\newrefformat{fig}{Fig.~\ref{#1}}
\newrefformat{plot}{Fig.~\ref{#1}}
\newrefformat{ex}{Example~\ref{#1}}
\newrefformat{list}{List~\ref{#1}}
\newrefformat{tab}{Table~\ref{#1}}
\newrefformat{enum}{Case~(\ref{#1}.)}
\newrefformat{def}{Definition~\ref{#1}}
\newcommand{\pref}[1]{\prettyref{#1}}

\usepackage[inline]{enumitem}
\SetEnumitemKey{:roman}{label=(\roman*)}
\SetEnumitemKey{:arabic}{label=(\arabic*)}

\usepackage{circledsteps}
\usepackage{lipsum}

\usepackage{tikz}
\newcommand*\circled[1]{\tikz[baseline=(char.base)]{
            \node[shape=circle,draw,inner sep=1.5pt] (char) {#1};}}

\usepackage{acro}
\makeatletter
\@ifpackagelater{acro}{2017/02/17}{\@ifpackagelater{acro}{2019/09/23}{}{
    \let\IDeclareAcronym\DeclareAcronym
    \renewcommand{\DeclareAcronym}[2]{%
        \IDeclareAcronym{#1}{%
        #2,foreign-plural={}
        }
    }
}}{}

\usepackage{siunitx}
\usepackage{xfp}
\DeclareSIUnit{\noop}{\kern 0pt}
\newcount\tmpnum

\def\storedataA#1{\advance\tmpnum by1
    \ifx\end#1\else
    \expandafter\def\csname data\tmp\the\tmpnum\endcsname{#1}%
    \expandafter\storedataA\fi
}

\def\assert#1{\ifthenelse{#1}{}{\errmessage{ASSERT FAIL}}}
\def\getarrdata[#1]#2{\ifcsname data#2#1\endcsname\csname data#2#1\endcsname\else\errmessage{UNSET #2#1}\fi}

\def\getarr[#1]#2{\getarrdata[#1]{#2}}

\def\roundprefixprefix[#1]#2{\SI[scientific-notation = engineering, exponent-to-prefix = true, round-mode=places,round-precision=#1]{#2}{\noop}}

\def\roundprefix[#1]#2{\ifthenelse {#2 < 1000}{#2\,\,~}{\roundprefixprefix[#1]{#2}}}

\def\round[#1]#2{\SI[round-mode=places,round-precision=#1,round-integer-to-decimal]{#2}{\noop}}

\def\rgetarr[#1]#2{\ifthenelse{\equal{\getarr[#1]{#2}}{ }}{}{\roundprefix[2]{\getarr[#1]{#2}}}}
\def\ngetarr[#1]#2{\ifthenelse{\equal{\getarr[#1]{#2}}{ }}{}{\SI[group-separator={\,}, group-minimum-digits=4]{\getarr[#1]{#2}}{\noop}}}

\def\calcpercprec#1#2#3{$\sim$\SI[round-mode=places,round-precision=#3]{\fpeval{((#1)*1.0)/((#2)*1.0)*100}}{\percent}}

\def\calcpercprec[#1]#2#3{$\sim$\SI[round-mode=places,round-precision=#1]{\fpeval{((#2)*1.0)/((#3)*1.0)*100}}{\percent}}
\def\calcpercprecnosim[#1]#2#3{\SI[round-mode=places,round-precision=#1]{\fpeval{((#2)*1.0)/((#3)*1.0)*100}}{\percent}}

\DeclareAcronym{RTT}{
  short        = RTT,
  long         = round-trip time
}
\DeclareAcronym{VEC}{
  short        = VEC,
  long         = Valid Edge Counter
}
\DeclareAcronym{TCP}{
  short        = TCP,
  long         = Transmission Control Protocol,
  first-style  = short
}
\DeclareAcronym{ECN}{
  short        = ECN,
  long         = Explicit Congestion Notification,
  first-style  = short
}

\begin{document}

\title[Does It Spin? On the Adoption and Use of QUIC’s Spin Bit]{Does It Spin? \\ On the Adoption and Use of QUIC’s Spin Bit}

\author{Ike Kunze}
\email{kunze@comsys.rwth-aachen.de}
\orcid{0000-0001-8609-800X}
\affiliation{%
  \institution{RWTH Aachen University}
  \country{Germany}
}
\author{Constantin Sander}
\email{sander@comsys.rwth-aachen.de}
\orcid{0009-0004-6627-1708}
\affiliation{%
  \institution{RWTH Aachen University}
  \country{Germany}
}
\author{Klaus Wehrle}
\email{wehrle@comsys.rwth-aachen.de}
\orcid{0000-0001-7252-4186}
\affiliation{%
  \institution{RWTH Aachen University}
  \country{Germany}
}
\renewcommand{\shortauthors}{Ike Kunze, Constantin Sander \& Klaus Wehrle}

\begin{abstract}  
Encrypted QUIC traffic complicates network management as traditional transport layer semantics can no longer be used for \acs{RTT} or packet loss measurements.
Addressing this challenge, QUIC includes an optional, carefully designed mechanism: the spin bit.
While its capabilities have already been studied in test settings, its real-world usefulness and adoption are unknown. 
In this paper, we thus investigate the spin bit’s deployment and utility on the web.

Analyzing our long-term measurements of more than \SI{200}{M} domains, we find that the spin bit is enabled on $\sim$10\% of those with QUIC support and for $\sim$\,50\% / 60\% of the underlying IPv4 / IPv6 hosts.
The support is mainly driven by medium-sized cloud providers while most hyperscalers do not implement it.
Assessing the utility of spin bit \acs{RTT} measurements, the theoretical issue of reordering does not significantly manifest in our study and the spin bit provides accurate estimates for around \SI{30.5}{\percent} of connections using the mechanism, but drastically overestimates the \acs{RTT} for another \SI{51.7}{\percent}.
Overall, we conclude that the spin bit, even though an optional feature, indeed sees use in the wild and is able to provide reasonable \acs{RTT} estimates for a solid share of QUIC connections, but requires solutions for making its measurements more robust.
\end{abstract}



\maketitle

\acresetall
\section{Introduction} 
\label{sec:introduction}

Network operators require visibility into their networks to locate and fix faults and performance bottlenecks.
Passive flow measurements are one commonly used source of information for this purpose:
they track traffic already passing through the network without adding additional overhead to the production traffic.
For example, \ac{TCP} semantics can be leveraged to estimate flow \acp{RTT}~\cite{Sengupta:Sigcomm2022:TCPRTT} or track reordering~\cite{Zheng:arXiv:TCPReordering} and packet loss~\cite{benkoe:GLOBECOM2002:PassiveE2ETCP}. 

Such passive methods are, however, challenged by the rising share of encrypted transport protocols.
QUIC, \eg denies passive observers access to the protocol header fields, making the mentioned TCP methods neither applicable nor adaptable to QUIC.
Instead, there is a push toward adding well-defined, explicitly measurable signals to protocols that can then be analyzed by passive observers~\cite{allman:CCR2017:Measurability,Cociglio:IETF2023:EFM}.
The QUIC spin bit is one such mechanism, enabling explicit \ac{RTT} measurements on otherwise encrypted traffic~\cite{RFC9000}.

While the addition of the spin bit is a promising sight for operators hoping to regain measurability of their networks, there is one significant hindrance: the spin bit is only an optional feature, \ie QUIC implementations do not have to support it.
Inspecting the publicly available QUIC stacks~\cite{QUICImplOverview} already reveals that only few implement the spin bit.
Furthermore, the spin bit has mainly been studied in test settings~\cite{devaere:IMC2018:ThreeBits,Bulgarella:ANRW2019:DelayBit}.
Whether or not network operators can rely on the spin bit for accurate \ac{RTT} estimation in the wild is thus an important open question and there is little publicly known information on these practical issues.

In this paper, we, thus, set out to investigate the adoption and accuracy of the spin bit in the wild.
Performing long-term, large-scale measurements of more than \SI{200}{M} domains, including many popular targets of toplists, we study the host support of the spin bit on the web, finding that it is active on $\sim$10~\% of domains that support QUIC and mostly driven by small- to medium-sized providers.
Judging the utility of spin bit-based measurements by comparing them to QUIC’s built-in \ac{RTT} estimates, we observe that the spin bit generally overestimates the real \ac{RTT} as expected, but still provides accurate measurements for around \SI{30.5}{\percent} of connections using the mechanism. 
Overall, we contribute the following:
\begin{itemize}
  \item We study the QUIC spin bit adoption in the web using repeated measurements of more than \SI{200}{M} domains.
  \item We find that around 10\,\% of domains with QUIC support also support the spin bit with adoption mainly driven by small to medium cloud providers.
  \item Assessing its measurement accuracy, we observe that the spin bit mostly overestimates the real \ac{RTT} and yields sensible estimates for a third of the studied cases.
\end{itemize}

\section{QUIC and Its Spin Bit} 
\label{sec:background}

The spin bit allows passive network observers to determine flow \acp{RTT} and can be seen as an answer to Allman\etal’s call for adding explicit measurement mechanisms into protocols~\cite{allman:CCR2017:Measurability}.
It addresses a need of network operators who, with the rising share of QUIC~\cite{rueth:PAM2018:QuicFirstLook,Zirngibl:IMC2021:Over9000}, increasingly lose visibility into the performance of their networks as established measurements relying on implicit protocol semantics, \eg exposed by \ac{TCP}, can no longer be used.

\begin{figure}[t]
  \centering
  \includegraphics[width=0.95\columnwidth]{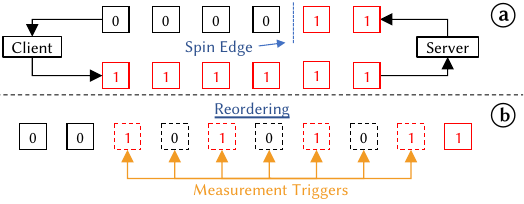}
  \caption{\Circled{a}: The client of a connection always \emph{flips} the spin bit; the server \emph{reflects} it. \Circled{b}: Reordering near the spin edges can cause faulty measurements.}
  \label{fig:spinbit}
\end{figure}

\subsection{The QUIC Spin Bit} 
\label{sub:the_quic_spin_bit}

The spin bit is an optional, version-dependent feature of QUIC version 1 that uses one bit in short header packets~\cite{RFC9000}. 
In short, it modulates a square wave onto a QUIC flow with a wavelength equal to the flow’s \ac{RTT} measurable by observers.

\afblock{Mechanism.}
The client starts the signal by transmitting packets with a value of $0$.
As visualized in \pref{fig:spinbit} \circled{a}, the server reflects the value it has received, setting the value on outgoing packets to the value seen on the latest incoming packet with the highest packet number.
In contrast, the client spins the bit, \ie it inverts the latest value.
Observers can estimate the \ac{RTT} by measuring the time between two consecutive spin bit flips ("spin edges").

\afblock{Manageability considerations.}
As the spin bit needs support from both peers of a connection, either endpoint can unilaterally decide to disable the mechanism. 
RFC~9000~\cite{RFC9000} further mandates that even end-points actively using the spin bit "MUST" disable it on at least one in every 16 connections to ensure that disabled spin bits appear regularly in Internet traffic.
One major reason for this decision is that some endpoints may want to disable the spin bit to prevent \ac{RTT} measurements, and this decision should not be cause for any mistreatment by the network~\cite{RFC9312}.
Whether these guidelines are indeed followed in practice is unknown.

\afblock{Disabling the spin bit.}
RFCs~9000 and~9312~\cite{RFC9000,RFC9312} recommend that end-points disable the bit via greasing, \ie setting it randomly on a per-packet or per-connection ID basis.
They can further set it arbitrarily, meaning that using the same value on all connections is also possible.
Hence, despite the recommendations, it is unknown which form of disabling is commonly chosen on the web.

\afblock{Accuracy considerations.}
Spin bit \ac{RTT} measurements are vulnerable to different causes of inaccuracy.
On the one hand, the spin bit always includes end-host delays, \eg the time needed to process a request, which might cause inflated \ac{RTT} estimates.
On the other hand, the measurements can be disturbed by packet reordering as is visualized in \pref{fig:spinbit} \circled{b}.
More specifically, reordering around spin edges can introduce ultra-short spin cycles and, hence, very short \ac{RTT} estimates.
While a solution addressing the first issue was originally proposed in the form of the \ac{VEC}~\cite{devaere:IMC2018:ThreeBits}, it has not found its way into the standard.
Addressing the second issue, RFC~9312~\cite{RFC9312} proposes the use of heuristics, but the usefulness and necessity of these heuristics has not yet been tested on a larger scale in academic work.

\subsection{Related Work on QUIC and the Spin Bit} 
\label{sub:works_on_quic_and_the_spin_bit}

Studies by Rüth\etal~\cite{rueth:PAM2018:QuicFirstLook} in 2018 and Zirngibl\etal~\cite{Zirngibl:IMC2021:Over9000} in 2021 show an increasing deployment of QUIC on the Internet and in the web.
However, neither study investigates the use of the spin bit, which is the only optional feature of QUIC.

Academic work on the spin bit goes back to De Vaere\etal’s three-bit mechanism~\cite{devaere:IMC2018:ThreeBits}, which, in addition to the spin bit, included the \ac{VEC} designed to provide more stable measurements.
The authors study the capabilities of the spin bit subject to packet loss and reordering, and compare it to traditional measurements using the \ac{TCP} timestamp option.
While their results show that the spin bit is a useful mechanism comparable to \ac{TCP} timestamps, the experiments are small-scale and do not consider the validity of spin bit measurements in larger deployment scenarios.
Bulgarella\etal~\cite{Bulgarella:ANRW2019:DelayBit} confirm the general findings, but again focus on small-scale testbed experiments.
Kunze\etal~\cite{Kunze:EPIQ2021:SpinTracker} show that the spin bit observer logic can be implemented on P4 hardware switches and that the heuristics of RFC~9312 can filter faulty \ac{RTT} measurements to some extent.

In summary, QUIC deployment numbers are increasing~\cite{rueth:PAM2018:QuicFirstLook,Zirngibl:IMC2021:Over9000} and the spin bit has been shown to provide accurate latency measurements in small-scale and testbed experiments~\cite{devaere:IMC2018:ThreeBits,Bulgarella:ANRW2019:DelayBit}.
However, its adoption in the wild, real-life usage patterns, \eg regarding the guidelines defined in RFCs~9000 and 9312~\cite{RFC9000,RFC9312}, and the practical measurement utility, are still unknown.

\section{Study Design} 
\label{sec:study_design}

The adoption and usefulness of the spin bit in the wild are unknown.
Hence, to shed light on this topic, we perform a large-scale Internet measurement campaign that focuses on the web.

\afblock{Approach.}
We first identify suitable targets for our study via domain top- and zonelists (\pref{sub:determining_target_hosts}).
We then conduct HTTP/3 requests via QUIC from our university network to these targets and our adapted webclient stores relevant connection information in the qlog~\cite{Marx:IETF:QLOG} format (\pref{sub:querying_the_target_population}).
Using these logs, we assess the spin bit adoption and its usefulness for \ac{RTT} measurements (\pref{sub:studying_the_target_population}).

\subsection{Determining Target Hosts} 
\label{sub:determining_target_hosts}
We assemble our target population using multiple datasets.

\subsubsection{Domain Toplists}
Relying on different toplists, we first gather a target set that focuses on popular domains:
\begin{enumerate*}[:roman]
\item Alexa Top 1M\footnote{The Alexa list is deprecated since 05-2022; it stopped changing in 02-2023}~\cite{Alexa},
\item Cisco Umbrella~\cite{Umbrella},
\item Majestic Million~\cite{Majestic}, and
\item the Tranco Research List~\cite{LePochat:NDSS19:TRANCO}.
\end{enumerate*}
We further deduplicate their entries.

\subsubsection{Domain Zonelists} 
\label{ssub:domain_zonelists}
For a broader assessment and to also include less popular domains, we enrich our target population with domains from zone files accessible via the ICANN Centralized Zone Data Service (CZDS).
For our measurements in calendar week (CW) 20, 2023, we use $1140$ individual zones, including various gTLDs such as .com, .net, and .org.
Combined, the top- and zonelists amount to a total of around \SI{219}{M} domains for CW 20, 2023.

\subsection{Querying the Target Population} 
\label{sub:querying_the_target_population}
We perform HTTP/3 requests to retrieve the landing webpages of the target population.

\subsubsection{Tooling} 
\label{ssub:tooling}
We use an adapted version of zgrab2~\cite{zgrab2}, a golang-based banner grabber, as the general framework for our measurements.
Underneath, we add \emph{quic-go}~\cite{quicgo}, which supports QUICv1, and extend it for draft versions 27, 29, 32, and 34, and the spin bit.
We prepend each domain with "www" for targeting websites, as is common practice~\cite{Zimmermann:Networking17:HTTPPush,Rueth:IMC18:CryptoMining}, and capture quic-go's qlog~\cite{Marx:IETF:QLOG} information (which we extended with the spin bit state) for each established connection.
Note that there can be more than a single connection for one domain, \eg due to redirects; to limit the impact of our measurements, we only follow up to 3 redirects.
Given our web-focused view, we further assume that the queried domains provide websites.
Hence, we cannot assess the behavior of domains not responding to HTTP/3.

\subsubsection{Vantage point and measurement schedule} 
\label{ssub:measurement_schedule}
We conduct our measurements from within the RWTH Aachen University network which connects to the well-connected German research network (DFN), \eg peering at DE-CIX with many ASes.
In general, we perform weekly measurements using IPv4.
We further perform measurements using IPv6 in selected weeks with the same methodology to allow for comparisons between IPv4 and IPv6 while still reducing the load of our measurements.
Our zonelist measurements start on Wednesdays and run through Fridays while the toplist measurements start on Fridays and finish by Saturdays.

\subsection{Studying the Target Population} 
\label{sub:studying_the_target_population}
We assess the behavior of the target population by systematically analyzing the collected qlog data.

\afblock{QUIC support.}
We first identify hosts supporting QUIC by checking whether connections could be established with remote endpoints.
For this, we check whether the endpoints answer to QUIC packets and whether the actual initialization process finishes.

\afblock{Spin bit support.}
For hosts with QUIC support, we focus on the received packets from the qlog and extract
\begin{enumerate*}[:roman]
  \item the spin bit state,
  \item the QUIC packet number, and
  \item the corresponding timestamp.
\end{enumerate*}
We identify connections that see spin bit flips, \ie values of both $0$ and $1$, as candidates that might support the spin bit.

\afblock{RTT assessment and grease filter.}
For the identified candidate connections, we use the spin bit mechanism to calculate the \acp{RTT} based on the received order of the packets.
We then compare these \acp{RTT} to estimates provided by the QUIC stack which relies on richer information.
In essence, the QUIC stack measures the time until a specific packet is acknowledged and additionally factors in processing delays as reported by the other host.
Based on these two kinds of \ac{RTT} values, we filter out connections that presumably grease the spin bit as soon as one spin bit \ac{RTT} estimate is smaller than the minimum of all QUIC client \ac{RTT} estimates.

\afblock{Impact of reordering.}
Finally, we also assess the impact of reordering on the spin bit measurements.
For this, we correct the ordering of the received packets based on the packet numbers and repeat the \ac{RTT} calculations.

\section{Spin Bit Use in the Wild} 
\label{sec:spin_bit_use_in_the_wild}

We characterize the use of the spin bit from diverse perspectives, aiming to identify usage patterns, drivers of its deployment, and its utility in the wild.
We base our assessment on long-term, large-scale IPv4 measurements (\cf \pref{sec:study_design}).
In this section, we focus on the general spin bit use before we analyze its \ac{RTT} measurement utility in \pref{sec:spin_bit_accuracy_in_the_wild}. 

\subsection{Dataset and Spin Bit Use Overview} 
\label{sub:spin_bit_use_overview}
We first give a broad overview on general statistics describing our measurement data and the use of the spin bit.
\pref{tab:overall_table_v4} shows the results of our latest IPv4 measurements in CW 20, 2023 for the \emph{Toplists}, all \emph{CZDS} lists, and a focused view on the .com, .net, and .org zones (\emph{com/net/org}).
General observations identified for CW 20 are also representative for our longitudinal measurements.  

\begin{table}
  \centering
  \setlength{\tabcolsep}{0.3em}
  \small
  \begin{tabularx}{210pt}{crrrrr}
    \toprule
    & & Total & \multicolumn{1}{c}{Resolved} & \multicolumn{1}{c}{QUIC} & \multicolumn{1}{c}{Spin} \\
    \midrule
    \multirow{2}{*}{Toplists} & \#Domains &

    \roundprefix[2]{2732702} &
    \roundprefix[2]{1937701} &
    \roundprefix[2]{547107} &
    \calcpercprecnosim[1]{37768}{547107} \\
    & \#IPs &
     &
    \roundprefix[2]{774832} &
    \roundprefix[2]{118544} &
    \calcpercprecnosim[1]{18011}{118544} \\

    \midrule
    \multirow{2}{*}{CZDS} & \#Domains &

    \roundprefix[2]{216520521} &
    \roundprefix[2]{183735238} &
    \roundprefix[2]{22205271} &
    \calcpercprecnosim[1]{2257938}{22205271} \\
    & \#IPs &
     &
    \roundprefix[2]{10271558} &
    \roundprefix[2]{259766} &
    \calcpercprecnosim[1]{117639}{259766} \\

    \midrule

    \multirow{2}{*}{com/net/org} & \#Domains &

    \roundprefix[2]{183047638} &
    \roundprefix[2]{158891771} &
    \roundprefix[2]{18415242} &
    \calcpercprecnosim[1]{2047280}{18415242} \\

    & \#IPs &
     &
    \roundprefix[2]{9203681} &
    \roundprefix[2]{242877} &
    \calcpercprecnosim[1]{112644}{242877} \\

    \bottomrule
  \end{tabularx}
  \caption{Overview of our IPv4 results for CW 20, 2023.}
  \label{tab:overall_table_v4}
\end{table}

\afblock{Domain view.}
The respective first rows show the overall number of domains contained in the different domain lists (\emph{Total}), the number of resolved domains (\emph{Resolved}), the number of domains with at least one QUIC connection (\emph{QUIC}), and the percentage of QUIC-enabled domains that are \emph{potentially} spin bit enabled, \ie see spin bit activity on at least one connection (\emph{Spin}). 
Looking at the toplists, we can see some spin bit activity for \SI{6.9}{\percent} of the corresponding QUIC connections.
In comparison, there is a bit higher activity for CZDS (\SI{10.2}{\percent}) and, especially, com/net/org (\SI{11.1}{\percent}).
Hence, there is small support for the spin bit, but it is not only focused on topdomains often provided by large hyperscalers.

\afblock{IP view.}
Drilling deeper into this consideration, we next study the individual IPs that are contacted for each domain to assess the host support for QUIC and the spin bit.
Here, we consider all IPs to which there is at least one connection with spin bit activity.
Taking a broad look at the results, we can see that the \SI{547}{k} QUIC-enabled toplist domains are served by $\sim$\SI{119}{k} IPs, \ie a ratio of \SI{21.7}{\percent}, while the corresponding ratios are smaller for CZDS (\SI{1.2}{\percent}) and com/net/org (\SI{1.3}{\percent}).
Surprisingly, almost \SI{50}{\percent} of the IPs serving CZDS and com/net/org show spin bit activity while only \SI{15.2}{\percent} of the toplist IPs show support.
This observation supports our initial conjecture that the main drivers of the spin bit might not be hyperscalers, but instead smaller providers.

\takeaway{About \SI{10}{\percent} of CZDS domains and \SI{50}{\percent} of IPs serving these domains show spin bit activity while there is significantly less support in the toplists of less than \SI{20}{\percent} of IPs.}


\subsection{Drivers of the Spin Bit} 
\label{sub:drivers_of_the_spin_bit}

Following our intuition that smaller providers might be at the forefront of spin bit support, we next aim to identify the main drivers of the spin bit.
For this, we inspect 
\begin{enumerate*}[:roman]
  \item the organizational affiliations of the supporting IPs, and
  \item the webservers serving these connections for com/net/org.
\end{enumerate*}

\afblock{Organizational affiliations.}
Analyzing the affiliation of all connections, we first map each IP to its corresponding ASN using BGP data of RIPE’s RIS archive~\cite{ripe:asn} and then lookup the corresponding organizations using CAIDA’s as2org dataset~\cite{caida:asn:org}. 

\begin{table}
  \centering
  \setlength{\tabcolsep}{0.5em}
  \small
  \begin{tabularx}{210pt}{rrlrrr}
    \toprule
    
    $\uparrow$ & \multicolumn{1}{c}{Total \#} & AS Organization & Spin \# & Spin \% & Spin $\uparrow$\\
    \midrule
    
    1     &
    \roundprefix[2]{11482201} &
    Cloudflare   &
    \roundprefix[2]{0} &
    \calcpercprecnosim[1]{0}{11482201} &
    0 \\
    
    2     &
    \roundprefix[2]{6160065} &
    Google   &
    \roundprefix[2]{6867} &
    \calcpercprecnosim[1]{6867}{6160065} &
    54 \\
    
    3     &
    \roundprefix[2]{1546788} &
    Hostinger   &
    \roundprefix[2]{802585} &
    \calcpercprecnosim[1]{802585}{1546788} &
    1 \\
    
    4     &
    \roundprefix[2]{326230} &
    Fastly   &
    \roundprefix[2]{0} &
    \calcpercprecnosim[1]{0}{326230} &
    0 \\
    
    5     &
    \roundprefix[2]{219249} &
    OVH SAS   &
    \roundprefix[2]{132395} &
    \calcpercprecnosim[1]{132395}{219249} &
    2 \\
    
    \midrule
    
    6     &
    \roundprefix[2]{218206} &
    A2 Hosting   &
    \roundprefix[2]{129577} &
    \calcpercprecnosim[1]{129577}{218206} &
    3 \\
    
    7     &
    \roundprefix[2]{173503} &
    SingleHop   &
    \roundprefix[2]{102527} &
    \calcpercprecnosim[1]{102527}{173503} &
    4 \\
    
    8 &
    \roundprefix[2]{148705} &
    Server Central &
    \roundprefix[2]{100518} &
    \calcpercprecnosim[1]{100518}{148705} &
    5 \\

    \midrule
    
    &
    \roundprefix[2]{2519770} &
    <other> &
    \roundprefix[2]{1342065} &
    \calcpercprecnosim[1]{1342065}{2519770} &
     \\ 
    
    \bottomrule
  \end{tabularx}
  \caption{QUIC connections and spin bit activity resolved for the corresponding AS organization for com/net/org and the IPv4 results of CW 20, 2023.}
  \label{tab:spin_per_asn_comnetorg}
\end{table}

\afblock{AS results.}
\pref{tab:spin_per_asn_comnetorg} shows the overall number of connections (\emph{Total~\#}) and the corresponding overall rank ($\uparrow$) for the top 5 organizations in terms of connection share.
We further add the number and percentage of connections with some spin bit activity (\emph{Spin~\#}/\emph{Spin~\%}) and the top 5 organizations in terms of absolute spin bit support as indicated by their rank (Spin$~\uparrow$).
As can be seen, the two organizations with the most connections, being responsible for more than \SI{75}{\%} of QUIC connections observable from our vantage point, show no (Cloudflare) or very little support (Google, rank 54) for the spin bit.
Instead, medium-sized hosters that also have a relatively high share of QUIC connections represent most of the spin bit support, each having spin activity on more than \SI{50}{\%} of their connections.
Hostinger has the largest absolute share with \SI{802.59}{k} connections with spin bit support.
Interestingly, \SI{53.3}{\percent} of the remaining \SI{2.52}{M} connections that do not belong to the top ASes show spin bit support, indicating that there is a broad base of support. 

Confirming our previous intuition, we can conclude that the spin bit is mainly in use by smaller organizations. 
This stands in contrast to general trends for QUIC for which Zirngibl\etal~\cite{Zirngibl:IMC2021:Over9000} find hyperscalers to be the main drivers.

\afblock{Webserver support.}
Given the broad spin bit support among many ASes, we next dig deeper into which servers support the spin bit.
For this, we look for HTTP server information, focusing on those connections where we can unambiguously match this information with the qlog information.

\afblock{Webserver results.}
By far the most connections reach \emph{LiteSpeed} webservers, making up more than \SI{80}{\%} of all connections (not shown), while another \SI{7}{\%} are served by \emph{imunify360-webshield}, which we also suspect to build upon \emph{LiteSpeed}.
Hence, we conclude that the overwhelming share of spin bit support is provided by LiteSpeed deployments, \ie likely a single stack.

\takeaway{Large drivers of QUIC, such as Cloudflare, and Google, do not support the spin bit.
Instead, smaller hosters, such as Hostinger, carry the bulk of spin bit support, although there is also large support among smaller ASes.
Most of this support can be traced back to LiteSpeed webservers.}

\begin{table}
  \centering
  \setlength{\tabcolsep}{0.3em}
  \small
  \begin{tabularx}{230pt}{crrrrr}
    \toprule
    & \multicolumn{1}{c}{All Zero} & \multicolumn{1}{c}{All One} & \multicolumn{1}{c}{Spin} & \multicolumn{1}{c}{Grease} \\
    \midrule
    Toplists  &
    \roundprefix[2]{507967} (\calcpercprecnosim[1]{507967}{547107}) &
    \roundprefix[2]{859} (\calcpercprecnosim[1]{859}{547107}) &
    \roundprefix[2]{37768}
    &
    \roundprefix[2]{58} (\calcpercprecnosim[1]{58}{547107})
    \\

    \midrule
    CZDS  &

    \roundprefix[2]{19849938} (\calcpercprecnosim[1]{19849938}{22205271}) &
    \roundprefix[2]{62375} (\calcpercprecnosim[1]{62375}{22205271}) &
    \roundprefix[2]{2257938}
    &
    \roundprefix[2]{5307} (\calcpercprecnosim[1]{5307}{22205271})
    \\

    \midrule
    com/net/org &

    \roundprefix[2]{16282445} (\calcpercprecnosim[1]{16282445}{18415242}) &
    \roundprefix[2]{53717} (\calcpercprecnosim[1]{53717}{18415242}) &
    \roundprefix[2]{2047280}
    &
    \roundprefix[2]{4653} (\calcpercprecnosim[1]{5307}{18415242})
    \\

    \bottomrule
  \end{tabularx}
  \caption{Spin behavior of all QUIC domains for our measurements in CW 20, 2023. Spin column corresponds to \pref{tab:overall_table_v4}.}
  \label{tab:status-codes}
\end{table}

\subsection{Spin Bit Configuration} 
\label{sub:spin_bit_configuration}

Having established the sources of spin bit support, we next dig deeper into how the spin bit is configured in practice.
RFCs~9000 and~9312~\cite{RFC9000,RFC9312} recommend that end hosts should disable the spin bit via greasing, \ie setting it to random values.
Greasing can be done on a per-packet or per-connection basis while using a fixed value is also possible.
We next assess which practices are used by analyzing how the spin bit is set in the wild.

\afblock{How is the spin bit disabled?}
\pref{tab:status-codes} shows how the different domains set the spin bit.
As can be seen, most domains that do not use the spin bit use a value of zero (\emph{All Zero}) while only few exclusively send a value of one (\emph{All One}).
Additionally, we only filter a small number of connections with our simplistic grease filter, indicating that \ac{RTT} underestimations do not occur frequently.

\afblock{Is the RFC "MUST" followed?}
Another critical aspect of RFC~9000 is that the spin bit "MUST" be disabled on one in 16 connections (or one in eight as per RFC~9312) to prevent ossification. 
We assess this configuration aspect by inspecting the behavior of domains over a longer period leveraging our longitudinal measurement data.
In particular, we first select $n$ measurement days spread across our \emph{measurement campaign} (CW 15, 2022 -- CW 20, 2023) and gather all domains that had the spin bit enabled on \emph{any} of these days.
We then select the domains to which we could establish a connection in \emph{every} week and count in how many weeks they showed spin bit activity.
We choose $n=12$ to include as many weeks as possible while still having a large number of considered domains.  

\begin{figure}
\centering
\includegraphics{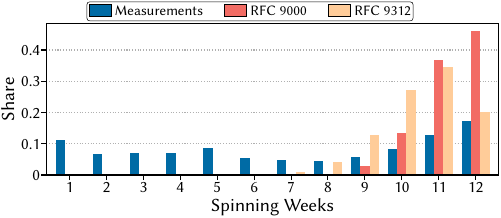}
\caption{Histogram showing the number of weeks in which the selected domains had the spin bit enabled. RFC values are computed using probability theory and reflect theoretical behavior when the spin bit is disabled in one in eight (RFC~9312) / 16 (RFC~9000) connnections.}
\label{fig:domain_evolution}
\end{figure}

Across all domain lists, we could identify a total of \SI{4.37}{M} domains that have spun at any time during the 12 selected weeks, out of which only  \SI{2.26}{M} domains had working connections in every week.
\pref{fig:domain_evolution} shows a relative histogram of the share of domains that had the spin bit enabled in the corresponding number of weeks as well as shares computed based on the RFC descriptions usinng probability theory.
As can be seen, slightly less than \SI{20}{\%} of domains spin the spin bit in all weeks while the shares for the other groups are around 5 to \SI{10}{\%} each.
Additionally, the domains spin less than would be allowed by the two interpretations of the specification, indicating that the guidelines on regularly disabling the spin bit are indeed followed.
We discuss an alternative methodology in \pref{sec:discussion}. 

\takeaway{
  Most hosts seem to opt for zeroing the spin bit.
  Additionally, greasing the spin bit on a per-packet basis, indicated by very short \ac{RTT} estimates, only occurs seldomly as indicated by small grease filter numbers.
  Finally, the long-term behavior of the spin-enabled domains shows that the RFC mandate seems to be generally followed.
}


\subsection{IPv6 and the Spin Bit} 
\label{sub:ipv6_and_the_spin_bit}
In addition to our IPv4 measurements, we have also conducted IPv6 measurements using the same methodology in some weeks of our measurement campaign to assess the state of the spin bit in the IPv6 space.
\pref{tab:ipv6-overview} shows the IPv6 results for CW 20, 2023.
While the overall number of domains with QUIC support is lower, we can see drastically more IPs supporting QUIC for CZDS and com/net/org, each also having a larger share of connections with spin bit support of more than \SI{60}{\%}.
In contrast, there is much lower spin bit support for the toplists of only \SI{2.3}{\percent} of domains and \SI{8.1}{\percent} of hosts.
Similar to the IPv4 results, LiteSpeed servers are responsible for around \SI{80}{\%} of all connections and Hostinger represents a large share of the spinning connections with \SI{91.5}{\percent}.
Overall, the IPv6 results draw a two-fold picture:
on the one hand, more modern IPv6 deployments seem to coincide with a higher spin bit support for CZDS and com/net/org.
On the other hand, the toplists, typically a driver for innovation, show a much worse support than with IPv4.

\takeaway{
  The spin bit support is broader for IPv6 compared to IPv4, yet worse when only focusing on the toplists.
  General trends regarding the main drivers of the spin bit are similar to IPv4.
}

\section{\ac{RTT} Measurement Accuracy} 
\label{sec:spin_bit_accuracy_in_the_wild}

While we have found that flows indeed use the spin bit on the web, the utility of the corresponding \ac{RTT} estimates is still unclear.
In particular, De Vaere\etal~\cite{devaere:IMC2018:ThreeBits} have identified different sources of inaccuracy when analyzing the performance of the spin bit in a testbed and a small set of real-world measurements; the impact of these issues on a larger scale and in practice is, however, unknown.
Hence, we now leverage our large-scale measurements to shed light on the practical accuracy of spin bit \ac{RTT} estimates.

\subsection{Assessment Methodology} 
\label{sub:assessment_methodology}
We base our assessment on the raw data of \emph{all} IPv4-based QUIC connections with spin bit activity throughout our \emph{entire measurement campaign} (\cf\,\pref{sub:spin_bit_configuration}), amounting to $\sim$\SI{86}{M} connections in total.
For each connection, we compare the mean of the spin bit estimates (\emph{spin}) to the mean of the QUIC stack estimates (\emph{QUIC}) using the following two methods.

\afblock{1) Absolute measurement accuracy.}
We assess the absolute accuracy of the spin bit by computing the difference between the two means: $abs=\emph{spin}-\emph{QUIC}$.

\afblock{2) Relative measurement accuracy.}
We assess the relative accuracy of the spin bit by computing the ratio of the means.
We always divide by the smaller of the two and multiply the ratio by $-1$ if $\emph{spin}<\emph{QUIC}$.
Negative values indicate an underestimation of the real \ac{RTT} while positive values represent an overestimation.

\afblock{Terminology.}
We assess the performance of spinning connections (\emph{Spin}) and connections filtered by our grease filter (\emph{Grease}).
We also distinguish results where we use the \emph{received} packet order (\emph{R}), potentially including reordering, and with packets \emph{sorted} according to their packet numbers (\emph{S}) to judge the impact of reordering.

\begin{table}
  \centering
  \setlength{\tabcolsep}{0.3em}
  \small
  \begin{tabularx}{210pt}{crrrrr}
    \toprule
    & & Total & \multicolumn{1}{c}{Resolved} & \multicolumn{1}{c}{QUIC} & \multicolumn{1}{c}{Spin} \\
    \midrule

    \multirow{2}{*}{Toplists} & \#Domains &

    \roundprefix[2]{2732702} &
    \roundprefix[2]{569516} &
    \roundprefix[2]{368331} &
    \calcpercprecnosim[1]{8511}{368331} \\

     & \#IPs &
     &

    \roundprefix[2]{166127} &
    \roundprefix[2]{94533} &
    \calcpercprecnosim[1]{7883}{94533} \\

    \midrule

    \multirow{2}{*}{CZDS} & \#Domains &

    \roundprefix[2]{216520521} &
    \roundprefix[2]{21467551} &
    \roundprefix[2]{9096258} &
    \calcpercprecnosim[1]{745370}{9096258} \\

     & \#IPs &
     &

    \roundprefix[2]{2115215} &
    \roundprefix[2]{1180320} &
    \calcpercprecnosim[1]{738791}{1180320} \\

    \midrule

    \multirow{2}{*}{com/net/org} & \#Domains &

    \roundprefix[2]{183047638} &
    \roundprefix[2]{17027333} &
    \roundprefix[2]{6626316} &
    \calcpercprecnosim[1]{675144}{6626316} \\

     & \#IPs &
     &

    \roundprefix[2]{1853223} &
    \roundprefix[2]{1041518} &
    \calcpercprecnosim[1]{662055}{1041518} \\

    \bottomrule
  \end{tabularx}

  \caption{Overview of our IPv6 results for CW 20, 2023.}
  \label{tab:ipv6-overview}
\end{table}

\subsection{Spin Bit Accuracy Results} 
\label{sub:spin_bit_accuracy}
\begin{figure}
\centering
\includegraphics{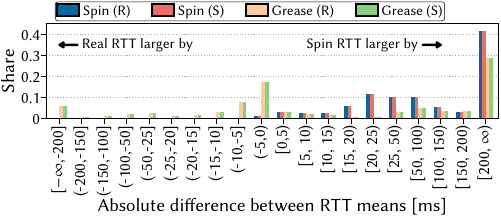}
\caption{Histogram of abs. difference between means of spin bit and QUIC estimate for all connections.}
\label{fig:rtt-mean-comsys-absolute}
\end{figure}

\begin{figure}
\centering
\includegraphics{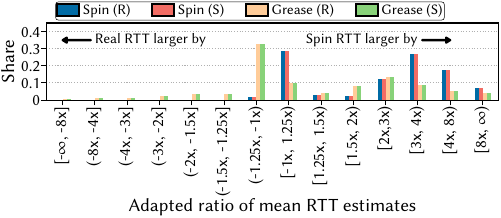}
\caption{Histogram of mapped ratio of the means of spin bit and QUIC estimate for all connections.}
\label{fig:rtt-mean-comsys-relative}
\end{figure}

\pref{fig:rtt-mean-comsys-absolute} shows a relative histogram of the mean differences for \emph{Spin} and \emph{Grease}, with (\emph{S}) and without (\emph{R}) correcting the packet order, while \pref{fig:rtt-mean-comsys-relative} shows the corresponding ratio.

\afblock{General usability.}
\SI{97.7}{\percent} of the Spin (R) results overestimate the real \ac{RTT} while there is only a small share of measurements underestimating it.
We can further observe that \SI{28.8}{\percent} of the connections have an absolute difference of $\le\SI{25}{\ms}$, yet \SI{41.3}{\percent} of the connections overestimate the real \ac{RTT} by more than \SI{200}{\ms}.

To assess the impact of these large absolute differences, we study the corresponding ratios in \pref{fig:rtt-mean-comsys-relative}.
As can be seen, the spin bit estimates are very close (less than 25\% difference) to the real \ac{RTT} for \SI{30.5}{\percent} of the spinning connections and \SI{36.0}{\percent} are within a factor of $2$.
However, a non-negligible portion of results (\SI{51.7}{\percent}) significantly overestimates the real \ac{RTT} by more than a factor of $3$.
Thus, while many estimates have an acceptable accuracy, the general tendency of the spin bit overestimating the real \ac{RTT} becomes apparent.

\afblock{Impact of reordering.}
Comparing the results of Spin (R) and Spin (S), we can observe that there is almost no perceivable difference between the two.
Going into more detail, we find differing results for only \SI{0.28}{\percent} of all connections with \SI{98.7}{\percent} of these differences having an absolute impact of less than \SI{1}{\ms}, slightly improving the measurement accuracy in \SI{93.1}{\percent} of the cases.
Hence, while reordering does occur, it does not seem to have a large-scale impact on the measurement accuracy, at least not for our vantage point, indicating that it might be more of a theoretical issue.

\afblock{Grease filtering.}
Inspecting the results of Grease (R/S), we suspect that our grease filter might create false positives.
While we do notice a larger share (\SI{46.0}{\percent}) of connections that underestimate the real \ac{RTT}, there are also many connections significantly overestimating it, although not as many as for Spin (R/S).
Furthermore, the relative impact of the underestimations seems to be rather small as \SI{62.5}{\percent} of the connections are within a factor of $2$.
We conjecture that the underestimations are not caused by greasing and suspect reordering to be one culprit, but only find \SI{7.9}{\percent} of the connections to be affected.
Thus, there seem to be yet unknown effects warranting further analysis which we enable by sharing our underlying spin bit measurement data set (\cf\pref{app:artifacts}).

\takeaway{The spin bit allows accurate \ac{RTT} estimates for many connections.
An even larger share sees drastic overestimations. 
While reordering has no pronounced impact on our results, unknown effects cause unexpected underestimation warranting further inspection. 
}

\section{Discussion} 
\label{sec:discussion}

Our measurement campaign shows that the spin bit is indeed used in the wild but that it is mainly driven by smaller providers.
We have further shed light on RFC compliance regarding disabling the spin bit and we find that many spin bit measurements are rather inaccurate.
In this section, we discuss possible reasons for and implications of our findings as well as limitations of our work.

\afblock{Reasons for disabling the spin bit.}
One deliberate reason for not using the spin bit are privacy concerns.
In particular, using the spin bit exposes the \ac{RTT} which can theoretically be used for locating users.
However, Trammell and Kühlewind~\cite{Trammell:PAM2018:LatencyPrivacy} have shown that these concerns are negligible.
A more pragmatic reason for not using the spin bit could be that QUIC stacks are still in the process of completing the mandated functionality.
Given that the spin bit does not provide an immediate benefit for the stack providers, implementing the optional spin bit might just not be a priority.

\afblock{Measurement accuracy.}
By design, spin bit \ac{RTT} estimates include end-host delays which are likely reasons for the identified large absolute differences (\cf\pref{sub:spin_bit_accuracy}).
These delays are most prominent at connection starts, which our approach focuses on, while measurements tend to stabilize over longer durations~\cite{Kunze:ANRW2021:LQRT}.
Hence, apart from assessing the usefulness of the spin bit for practical applications, such as network tomography~\cite{Coates:IEEESigProc2002:InetTomography}, it might be worthwhile to study the spin bit accuracy for longer connections.

\afblock{Limitations.}
Due its large footprint and other measurements conducted from our vantage point, our measurement campaign only performs weekly one-shot measurements without the ability to run additional measurements.
As a consequence, we use the results of consecutive weeks for assessing RFC 9000 and 9312 compliance.
Hence, our results might be affected by longer-term deployment changes as we note when studying the use of \ac{ECN} in QUIC based on the same dataset~\cite{Sander:IMC23:ECNQUIC}.
To study RFC compliance more accurately, one could first identify domains with an enabled spin bit in a large-scale measurement and then follow up with multiple measurements of a smaller target set, \eg querying them n=$16$ times to more accurately reflect RFC descriptions.

\section{Conclusion} 
\label{sec:conclusion}

The spin bit promises to provide network operators with visibility into the latency behavior of their networks in times of rising shares of QUIC traffic.
In this paper, we present the results of our measurement campaign investigating the practical use of the spin bit on more than \SI{200}{M} domains, including many popular targets.

We find that around \SI{10}{\%} of domains with QUIC support also use the spin bit, fueled by $\sim$50 (IPv4) to $\sim$60\,\% (IPv6) support of the underlying hosts.
The support is mainly driven by small to medium cloud providers, each using the spin bit on more than \SI{50}{\%} of their connections.
For connections with spin bit support, we find a high measurement accuracy of less than \SI{25}{\%} difference to the baseline for around \SI{30}{\%} of connections while \SI{50}{\%} of connections experience an overestimation by a factor of more than 3.

Concluding, we find a rather high adoption of the spin bit given the low support in QUIC stacks.
Our \ac{RTT} utility assessment further suggests that the spin bit provides sensible capabilities for network operators for a solid share of connections, but that estimates have to be used with caution and can benefit from further research, \eg studying the usefulness of filtering techniques described in RFC~9312~\cite{RFC9312}.
Overall, it remains to be seen if implementation support by hypergiants will follow naturally with growing maturity of the QUIC stacks or if it requires strong incentives, \eg in the form of immediate benefits enabled by the spin bit.

\clearpage

\section*{Acknowledgments}
This work has been funded by the German Research Foundation DFG under Grant No. WE 2935/20-1 (LEGATO).
We thank the anonymous reviewers and our shepherd Vaibhav Bajpai for their valuable feedback.
We further thank Leo Blöcher for maintaining our quic-go / zgrab2 changes and the network operators at RWTH Aachen University, especially Jens Hektor and Bernd Kohler.

\bibliographystyle{ACM-Reference-Format}
\balance
\bibliography{literature}

\appendix

\section{Ethics.} 
\label{app:ethical_considerations_}
We follow widely accepted ethical measures and measurement guidelines to minimize the impact of our study~\cite{Durumeric:UsenixSec2013:ZMap,Dittrich:USHS2012:MenloReport}.

\afblock{Information.}
We inform about our research by 
\begin{enumerate*}[:arabic]
  \item adding reverse DNS entries to our measurement IPs showing our research context,
  \item embedding our projectname as hint in every HTTP request, and
  \item hosting a dedicated website on our measurement IPs.
\end{enumerate*}
The website explains our research and how measurement targets can opt out of our study: via a single email or by blocking our measurement subnet. 
We quickly handle all (abuse) emails.

\afblock{Load reduction.}
We conduct our measurements from within our network using a dedicated IP of our measurement subnet.
We spread each measurement over multiple days to reduce the peak load on networks, end-hosts, and shared infrastructure, such as DNS, \eg only processing $1000$ domains per second. 
Furthermore, we maximize the utility of our measurements as distinct partial results are reused for different research projects (\eg \cite{Sander:IMC23:ECNQUIC}), \ie we only perform one set of measurements instead of several individual ones, significantly reducing the caused load.

\section{Artifacts}
\label{app:artifacts}
We release our changes to zgrab~\cite{COMSYS:Zgrab}, quic-go~\cite{COMSYS:QuicGo} and our raw measurement data of the toplists~\cite{Zgrab:Raw:toplists}, stripping unused information to limit the file size.
To enable future work on the spin bit, \eg regarding \ac{RTT} filtering mechanisms (\cf\pref{sub:spin_bit_accuracy}), we also add the extracted raw spin bit information for all domains as detailed in \pref{sub:studying_the_target_population} together with qlog baseline information and corresponding analysis scripts~\cite{Spinbit:Raw}.

\end{document}